\documentclass[pdflatex,sn-mathphys-num]{sn-jnl}

\usepackage{graphicx}%
\usepackage{multirow}%
\usepackage{amsmath,amssymb,amsfonts}%
\usepackage{amsthm}%
\usepackage{mathrsfs}%
\usepackage[title]{appendix}%
\usepackage{xcolor}%
\usepackage{textcomp}%
\usepackage{manyfoot}%
\usepackage{booktabs}%
\usepackage{algorithm}%
\usepackage{algorithmicx}%
\usepackage{algpseudocode}%
\usepackage{listings}%
\usepackage{subcaption}
\usepackage{graphicx}
\usepackage{amssymb}
\usepackage{array}
\usepackage{booktabs}
\usepackage{pifont}

\newcommand{\xmark}{\ding{55}}
\newcommand{\cmark}{\checkmark} 

\theoremstyle{thmstyleone}%
\theoremstyle{thmstyletwo}%

\theoremstyle{thmstylethree}%

\begin{document}

\title[Article Title]{An Improved Fault Diagnosis Strategy for Induction Motors Using Weighted Probability Ensemble Deep Learning}


\author*[1]{\fnm{Usman} \sur{Ali}}\email{usmanali@gift.edu.pk}



\author[1]{\fnm{Umer} \sur{Ramzan}}\email{umer.ramzan@gift.edu.pk}

\author[2]{\fnm{Waqas} \sur{Ali}}\email{waqas.ali@uet.edu.pk}





\abstract{Early detection of faults in induction motors is crucial for ensuring uninterrupted operations in industrial settings. Among the various fault types encountered in induction motors, bearing, rotor, and stator faults are the most prevalent. This paper introduces a Weighted Probability Ensemble Deep Learning (WPEDL) methodology, tailored for effectively diagnosing induction motor faults using high-dimensional data extracted from vibration and current features. The Short-Time Fourier Transform (STFT) is employed to extract features from both vibration and current signals. The performance of the WPEDL fault diagnosis method is compared against conventional deep learning models, demonstrating the superior efficacy of the proposed system. The multi-class fault diagnosis system based on WPEDL achieves high accuracies across different fault types: 99.05\% for bearing (vibrational signal), 99.10\%, and 99.50\% for rotor (current and vibration signal), and 99.60\%, and 99.52\% for stator faults (current and vibration signal) respectively. To evaluate the robustness of our multi-class classification decisions, tests have been conducted on a combined dataset of 52,000 STFT images encompassing all three faults. Our proposed model outperforms other models, achieving an accuracy of 98.89\%. The findings underscore the effectiveness and reliability of the WPEDL approach for early-stage fault diagnosis in IMs, offering promising insights for enhancing industrial operational efficiency and reliability.}

\keywords{Induction motor, Ensemble learning, Deep learning, Fault diagnosis, STFT spectral Images}



\maketitle

\section{Introduction}\label{sec1}
The Induction Motors (IMs) serve as integral electromechanical components within the industrial sector, primarily employed in the fields of production, energy generation, and transport due to their inexpensive and ruggedness\cite{ali2020towards}. In recent years, extensive studies have been undertaken in the area of fault identification and classification for IMs, underscoring their crucial role in diverse sectors\cite{b2}. Faults in IMs lead to prolonged downtimes, resulting in significant losses due to maintenance expenses and revenue reduction.These types of faults are classified as either electrical or mechanical.  Mechanical faults are associated with bearings and eccentricity and electrical faults primarily occur in the rotor and stator\cite{b3}. These faults can be measured by analyzing the IMs' current, voltage, and vibration signals. Typically, the accuracy of fault classification hinges on selecting the appropriate signal and employing data collection techniques that offer vital insights into the motor's condition.
\begin{figure}[http]
    \centering
    \includegraphics[width=0.8\textwidth]{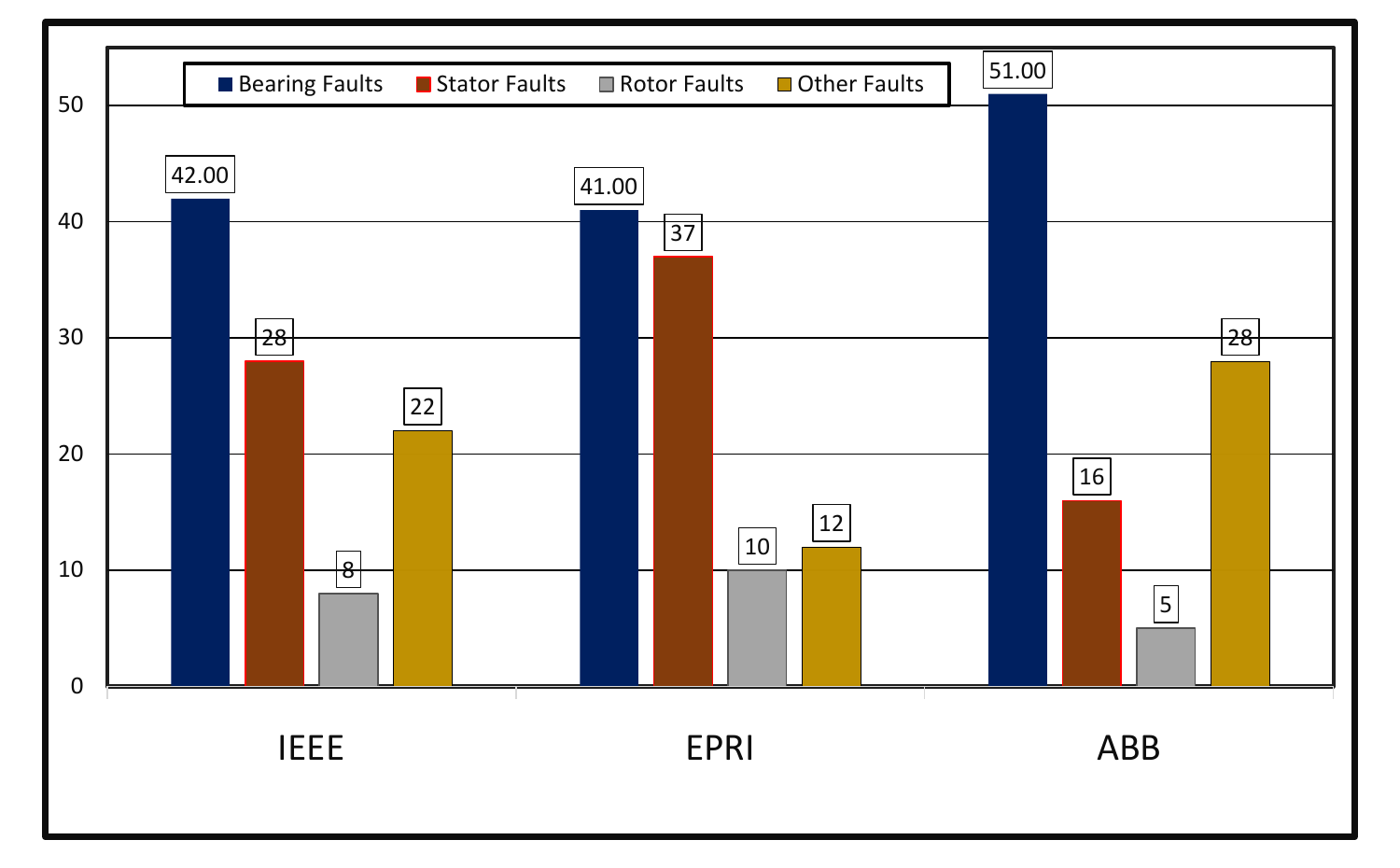} 
    \caption{Common Faults in IMs}
    \label{fig1}
\end{figure}
 Current monitoring and vibration signal measurements are predominantly employed for IMs to achieve precision due to their non-intrusive nature and resilience\cite{b4}. According to IEEE, EPRI\cite{b5}, and ABB\cite{b6} the distributions of common faults are shown in figure \ref{fig1}.

In recent years, multiple methodologies have been employed for detecting and categorizing bearing, stator, and BRB faults in IMs. These encompass thermal, induced voltage, variation of torque, vibration, and motor current signature analysis (MCSA)\cite{b7},\cite{b8}. Similarly, various machine learning (ML) and deep learning (DL) methods have been used to identify and classify these faults in IMs. These included logistic regression (LR), k-nearest Neighbors (KNN), support vector machines (SVM), decision trees (DT), ensemble learning (EL), artificial neural networks (ANN), convolution neural networks (CNN), and recurrent neural networks (RNN) \cite{ali2020towards}.

As researchers delve into the field of fault diagnosis in IMs, considerable progress has been achieved, as evidenced by the literature review discussed in \ref{sec2}. However, amidst the advancements, addressing a significant limitation prevalent in much of the prior research is essential. Specifically, many experiments conducted on existing techniques utilize relatively small datasets. This reliance poses a considerable challenge as it may result in overfitting concerns when implementing these techniques in industrial environments. Overfitting occurs when a model learns to perform well on the training data but fails to generalize to new, unseen data. Addressing this challenge is crucial for ensuring the practical applicability and effectiveness of these techniques in industrial environments, where data volumes are typically larger and more complex.
Furthermore, a distinct gap in the existing body of research pertains to the absence of an efficient technique capable of accurately identifying and classifying multi-class faults, particularly those associated with bearing, stator, and rotor malfunctions in IMs. This deficiency becomes particularly pronounced when confronted with high-dimensional datasets, posing a considerable obstacle to the precise diagnosis of faults in practical industrial applications.

In response to these identified limitations and gaps, this research work proposed an approach leveraging a deep learning-based ensemble method. This method is specifically tailored to handle high-dimensional datasets, enabling efficient identification and classification of diverse types of multi-class faults observed in IMs.

 The paper outlines the following objectives and contributions.   
\begin{itemize}
    \item A WPEDL mechanism is proposed for multi-class fault diagnosis of IMs, seamlessly integrating lightweight and other DL models. These models excel at learning the intricate high-dimensional features associated with faults observed in IMs.
    \item The STFT image processing technique is applied to extract time-frequency features, encompassing both transient and steady-state characteristics, from current and vibration signals.
    \item Utilizing a high-dimensional STFT spectral image dataset (as indicated in Table \ref{STFT Spectrograms in Each Class}) for training and testing to enhance generalization accuracy. The datasets for the rotor, stator, and bearing faults are accessible on IEEE data port\cite{data1}, Mendeley Data \cite{data2}, and Machinery Fault Database \cite{data3}. 
   \item In the results section, a detailed experimental analysis is provided, including both individual and combined datasets of faults in IMs. This analysis employs various evaluation metrics to ensure a comprehensive assessment. 
    \end{itemize}
The subsequent sections of this paper are outlined as follows: Section \ref{sec2} discusses the relevant work.  Section \ref{sec3} delves into the data acquisition process, while Section \ref{sec4} outlines the proposed methodology. Section \ref{sec5} is dedicated to the discussion and presentation of experimental results. Finally, Section \ref{sec6} concludes our work.
\section{Related Work}\label{sec2}

Researchers have applied various SP, ML, and DL techniques to address the fault classifications of IMs. Bearings are recognized as comprising inner and outer races that provide support to motor components. The operational integrity of these bearings is susceptible to issues such as wear and misalignment, which can profoundly impact their efficiency and functionality.
Zhang et al. \cite{b100} implemented a diagnostic approach for bearing faults using deep-CNN. Their method enhances fault diagnosis accuracy in noisy environments and across different workloads through innovative training methods and optimized network structures. However, the data augmentation techniques utilized in this approach might introduce additional noise in data.
Qian et al. \cite{b101} employed sparse filtering coupled with a high-order Kullback-Leibler technique for feature extraction. Subsequently, TL-based multi-class logistic regression has been utilized for feature classification, facilitating the integration of domain adaptation with health condition analysis of bearing faults.
Spyridon et al.\cite{b102} presented an approach to the diagnosis of rolling bearing faults by integrating an attention mechanism with a dense convolutional block.  Their method is designed to efficiently and accurately detect and identify bearing faults on the limited dataset by analyzing vibration signals.
Khorram et al. \cite{b103} employed the vibration signal from a gear bearing as input and applied a CNN to extract features and applied these features to a long short-term memory network for time series analysis and fault diagnosis.
EL has become increasingly popular within the realm of ML\cite{b104}-\cite{b105}. Alam et al. \cite{b106} introduced a neural network algorithm centered on dynamic EL. This algorithm dynamically modifies the model structure to accommodate various data types and features. Nonetheless, it could face challenges in handling extensive and intricate datasets. Webb et al. \cite{b107} presented a multi-strategy ensemble learning approach, which integrates diverse ensemble learning techniques to enhance performance and generalization ability. This approach demonstrates adaptability to various data types and task demands, while also exhibiting robustness and scalability.
In \cite{b108}, the authors utilized the EL technique where multiple ML models are combined to improve fault classification in IMs. By blending the strengths of random forest and extreme gradient boosting classifiers, the approach achieves higher accuracy and robustness in fault detection. One limitation of this study is that the experiments are conducted on a small-scale dataset, which may restrict the generalizability of the findings.
In \cite{b109}, the authors implemented an approach for fault diagnosis in industrial motors by blending statistical and DL features. It combines ensemble techniques, such as Extreme Gradient Boosting (XGBoost) with Particle Swarm Optimization (PSO), with a DNN for adaptive feature extraction. The ensemble method enhances classification accuracy and robustness across diverse datasets and noise levels, promising improved fault diagnosis in industrial contexts.

The stator of an induction motor consists of coils wound around iron cores, where inter-coil and inter-turn faults can occur, impacting motor performance and reliability.
In \cite{S1}, the authors proposed a new method for detecting stator inter-turn faults (SITFs) in induction motors using a 2D CNN. By analyzing fundamental frequency phasor magnitude (FPM) and 3rd harmonic components of stator currents, it achieves robust fault detection, validated experimentally on a 2.2 kW motor under various conditions. Both FPMs and SITFs contribute to fault detection and severity assessment, with FPMs effectively identifying the faulty phase.
In \cite{S2}, the author used a CNN architecture for detecting and classifying induction motor stator winding faults, particularly inter-turn short circuits. While demonstrating promising results in real-time fault detection and classification accuracy on a specially designed setup with a 3 kW induction motor, the study acknowledges the limitation of having less available data for training.
In \cite{s3}, the researchers presented a method for early stator fault diagnosis in IMs. It uses wavelet denoising and statistical analysis to extract fault features from current signals, enhancing accuracy with an ensemble AdaBoost decision tree classifier. Experimental validation demonstrates robustness, achieving 98.48\% correctness.
In \cite{S4}, the researchers proposed an online approach for detecting inter-turn short-circuit failures, employing DWT and SVM classification on a motor with a variable frequency drive. However, the study utilized only a limited number of fault severities and three loads.
In \cite{S5}, the authors introduced a method for fault diagnosis in three-phased Permanent magnet synchronous motors using vibration and current signal fusion. Stator faults were induced and analyzed using experimental datasets, with AdaBoost utilized for classification. The fusion of vibration and current data achieved an accuracy of 90.7\%.

 Rotor, the rotating part, has isolated iron core prone to cracks due to various stresses and conditions. Defective bars generate sidebands at equal distances from the fundamental frequency. \cite{ali2023test}.   
Wagner et al. presented four distinct pattern identification approaches encapsulating four ML techniques i.e., SVM, KNN, multi-layer perceptron, and a fuzzy ARTMAP, utilized for the binary and multi-classification of BRB defects in inverter-fed IMs. For the binary and multi-class classification, the implemented techniques predicted the model accuracy at 90\% and 95\% respectively on 1274 experimental samples\cite{b17}.
In\cite{b18}, the author employed multi-level feature extraction techniques, including Discrete Wavelet Transform (DWT) and binary signature combined with nearest component analysis. These extracted features then applied to the SVM and KNN classifier algorithms, resulting in a commendable success rate of 99.8\%. However, a notable limitation of this method is the risk of overfitting, given that the model trained on a relatively small dataset.
In\cite{b19}, the authors employed gradient histograms to derive parameter weights from the three-phase current of the IMs. Subsequently, these measured features were applied to train a multi-layer ANN enabling the model to discern intricate patterns and relationships within the data. However, the model's performance assessment was conducted on a relatively limited dataset of 229 experimental samples, resulting in an accuracy of 95\%. Li et al. introduced a methodology based on  SVM to diagnose gearbox conditions. They successfully identified issues such as broken bars, missing teeth, and cracked gears, achieving an average accuracy at 91\%\cite{b20}.
Siyu et al. implemented an advanced diagnostic approach employing a pre-trained VGG-16 deep neural network to effectively identify gearbox and bearing faults in induction motors (IMs). To enhance the model's discriminative capabilities, the researchers applied wavelet transformation to the time series data, extracting crucial time-frequency features. These features were subsequently utilized to fine-tune the VGG-16 model, which underwent training with 6000 samples for gearbox faults and 5000 samples for bearing faults. Remarkably, the model demonstrated a high accuracy of 99.8\% \cite{b21}. 
Shafi et al. employed a method by implementing greedy-gradient graph-based semi-supervised learning to identify both binary and multi-class faults in IMs. The process involved utilizing ten data DWT windows, each comprising 9000 data samples and applying curve-fitting techniques to extract essential data features. Remarkably, their model demonstrated an impressive accuracy of 97\% \cite{b22}. 
In\cite{b24}, the researchers implemented three models—CNN, unidirectional LSTM, and bidirectional LSTM—to forecast rotor faults. The outcomes indicated that CNN exhibited superior performance compared to the other models.
Sajal et al. analyzed an open-source vibration dataset of broken bar faults in IMs using various CNN variants. They enhanced data representation with STFT-extracted features, leading to improved model performance. Impressively, the VGG-16 model achieved a 97\% accuracy rate.\cite{b25}. 
Kevin et al. utilized six distinct CNN-based architectures, including VGG16, Inception V4, NasNETMobile, ResNet152, and SENet154, to conduct a multi-class classification of induction motors (IMs). Among these architectures, the VGG-16 model demonstrated notable performance on the small experimental dataset, comprising 16,050 samples, achieving an impressive accuracy of 99.8\%, accurately predicting the class labels\cite{b23}.

\section{Dataset Description}\label{sec3}
In this research work, three distinct open-source datasets i.e., rotor \cite{data1}, stator \cite{data2}, and bearing \cite{data3} have been used to conduct the experiments.

\subsection{Rotor Dataset Description}
\begin{figure}[http]
    \centering
    \includegraphics[width=0.8\textwidth]{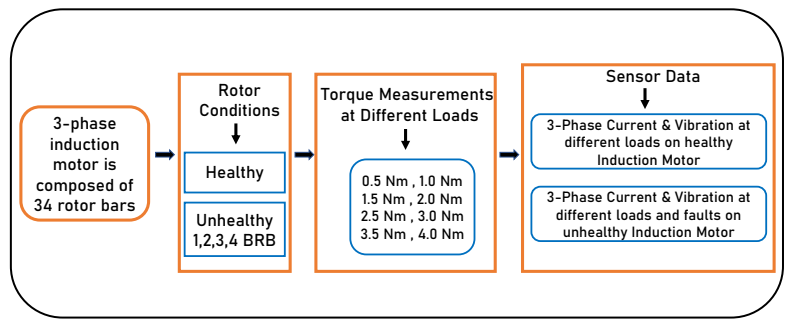} 
    \caption{Rotor Dataset Description Flow}
    \label{fig5}
\end{figure}

Figure \ref{fig5} shows the rotor dataset description flow. The rotor dataset is centered around a 1-horsepower IM operating at voltages of 220V/380V and drawing currents of 3.02A/1.75A.
It features 4 poles, operates at a frequency of 60 Hz, and runs at a speed of 1715 rpm. The rotor, of squirrel cage type, comprises 34 bars.
Testing encompasses load capacities on 12.5\%, 25\%, 37.5\%, 50\%, 62.5\%, 75\%, 87.5\%, and 100\%.
Electrical signal measurements are conducted precisely, utilizing alternating current probes with a capacity of up to 50ARMS and output voltage of 10 mV/A. 
For mechanical signal assessment, five axial accelerometers are employed. They possess a sensitivity of 10 mV/mm/s, a frequency range of 5 to 2000Hz, and stainless steel housing, enabling vibration measurements on both drive and non-drive ends in horizontal and vertical orientations.
Simultaneous sampling of signals occurs consistently over 18 seconds for each loading condition, across ten repetitions. This process captures both transient and steady-state phases of the induction motor. The data files are formatted in MATLAB (.mat) and contain information on four rotor classes for analysis: healthy, one, two, three, and four BRB faults.

\subsection{Stator Dataset Description}
\begin{figure}[http]
    \centering
    \includegraphics[width=0.82\textwidth]{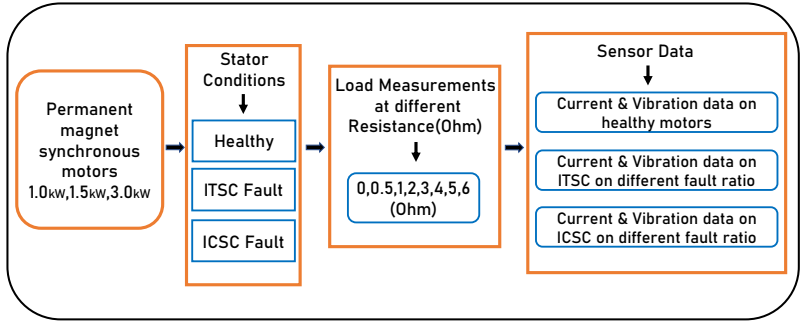} 
    \caption{ Stator Dataset Description Flow}
    \label{fig6}
\end{figure}

Figure \ref{fig6} shows the stator dataset description flow. The stator dataset includes vibration and current data from three PMSMs (1.0 kW, 1.5 kW, and 3.0 kW). 
Each motor has 16 stator faults: 8 inter-coil circuit faults and 8 inter-turn circuit faults. Motors run at 3000 RPM under a load condition of 15\% torque limit (1.5 Nm).  Vibration data, recorded using accelerometers, has a sampling frequency of 25.6 kHz for 120 seconds. Current data, measured with CT sensors, has a sampling frequency of 100 kHz for 120 seconds. Vibration amplitude is in g, and current is in A. The data files are formatted in technical data management streaming (.tdms) and contain information on three stator classes for analysis: healthy, inter-turn short circuit (ITSC) fault, and inter-coil short circuit (ICSC) fault.

\subsection{Bearing Dataset Description}
\begin{figure}[http]
    \centering
    \includegraphics[width=0.82\textwidth]{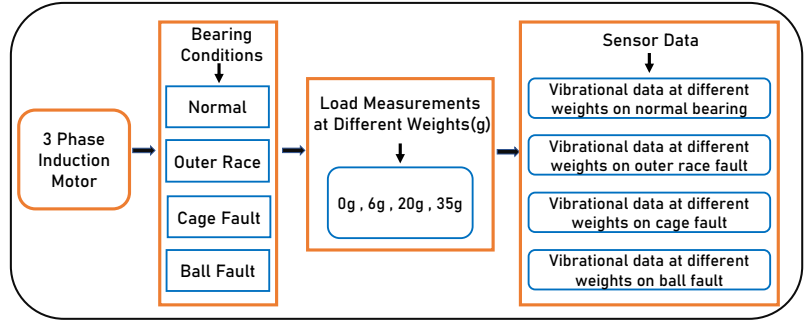} 
    \caption{Bearing Dataset Description Flow}
    \label{fig7}
\end{figure}

Figure \ref{fig7} shows the bearing dataset description flow. 
The bearing fault database comprises 1951 multivariate time series obtained from sensors installed on SpectraQuest's Machinery Fault Simulator (MFS) Alignment-Balance-Vibration (ABVT). These time series represent six distinct simulated states, including normal function and various fault conditions such as imbalance, horizontal and vertical misalignment, as well as inner and outer bearing faults.
The experimental bench specifications include a 1/4 hp motor with a frequency range of 700-3600 rpm, a system weight of 22 kg, and a rotor diameter of 15.24 cm. The distance between bearings is 390 mm, with eight balls having a diameter of 0.7145 cm and a cage diameter of 2.8519 cm.
For data acquisition, the system utilizes Industrial IMI Sensors, including three Model 601A01 accelerometers for radial, axial, and tangential directions, along with one Model 604B31 triaxial accelerometer. These sensors have a sensitivity of ±20\% (100 mV per g), a frequency range of 16-600000 CPM, and a measurement range of ±50 g.
Data acquisition is facilitated by two National Instruments NI 9234 modules, each equipped with four analog acquisition channels and a sample rate of 51.2 kHz.

\section{Proposed Methodology}\label{sec4}
\begin{figure}[http]
    \centering
    \includegraphics[width=1\textwidth]{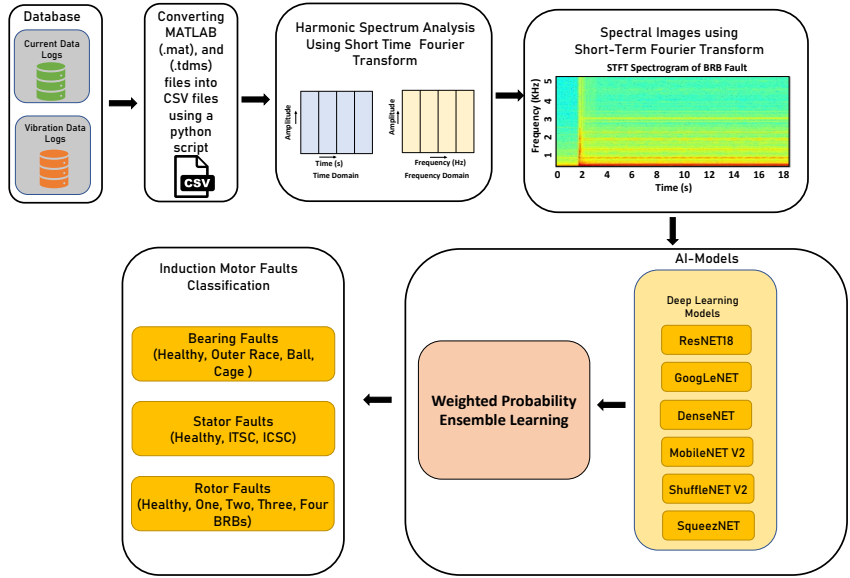} 
    \caption{General Workflow of the Proposed  Methodology}
    \label{fig8}
\end{figure}

The general workflow of the proposed working methodology is depicted in figure \ref{fig8}. Initially, data retrieval entails extracting information from the database, where .mat and .tdms format files undergo conversion to CSV files via a Python script. These CSV files are subsequently subjected to preprocessing, wherein statistical techniques are employed to normalize the data. Following preprocessing, STFT is utilized for visualizing the fundamental and sideband harmonic spectra inherent in the rotor, stator, and bearing datasets.

\begin{figure*}[http] 
    \centering
    \includegraphics[width=0.95\textwidth, height=0.95\textwidth]{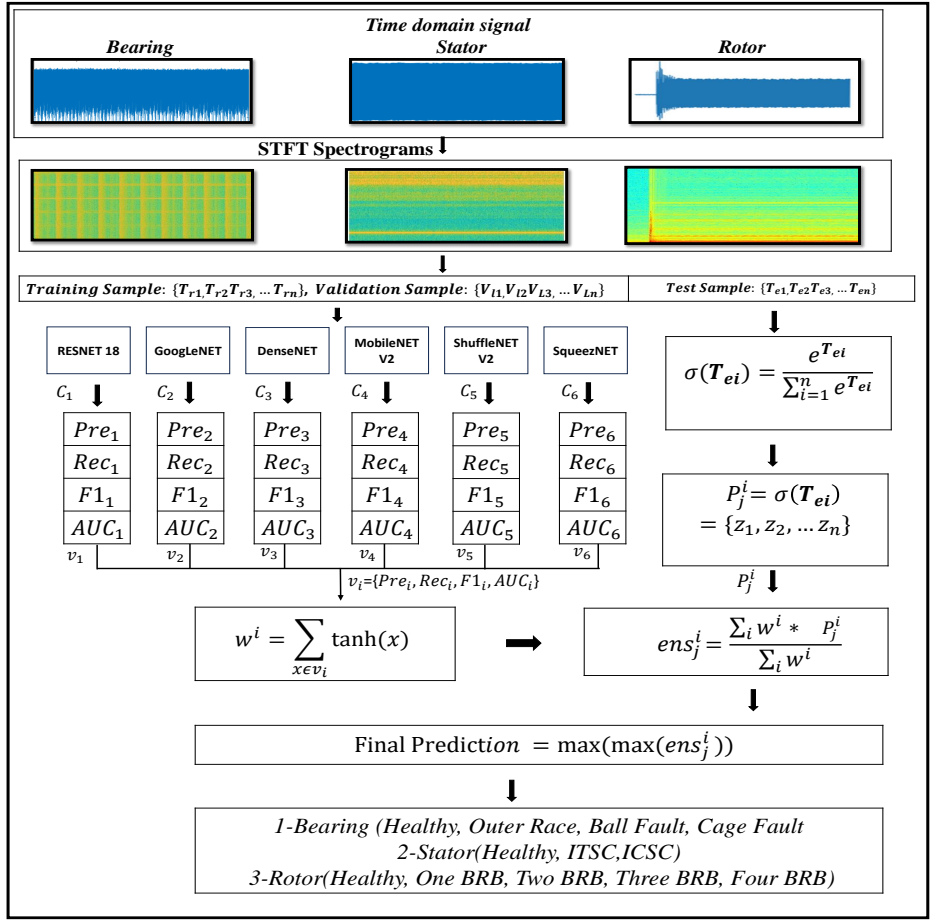} 
    \caption{Step-by-Step Working of WPEDL Approach}
    \label{fig9}
    \end{figure*}
    
Additionally, STFT offers a powerful method for analyzing time-varying signals by breaking them down into frequency components over short, overlapping time intervals. This process results in spectral images that provide valuable insights into the frequency content of the data across time.
The extracted RGB features from the STFT-transformed data serve as valuable inputs for the pre-training phase of various DL models. These models, including RESNET18, GoogleNet, DenseNET, MobileNETV2, ShuffleNETv2, and SqueezeNET, are chosen for their diverse architectures and capabilities in learning complex patterns from the data.

Furthermore, the weights obtained after fine-tuning these DL models on the specific fault classification task are saved for future use. The fine-tuning process entails refining the model's parameters to more accurately address the complexities inherent in fault detection within IMs. Finally, the weighted probability ensemble classification technique is applied, leveraging the saved model weights. This ensemble approach combines the predictions of multiple DL models, assigning different weights to each model based on its performance and reliability. The resulting ensemble classification enhances the overall fault diagnosis accuracy, enabling robust and effective identification of various multi-class faults in IMs.

Figure \ref{fig9} shows the detailed working methodology of our proposed WPEDL classification framework.

The proposed steps for diagnosing faults in an IM are summarized as follows:

\begin{table}[http]
  \centering
  \caption{STFT Spectrograms in Each Class}
  \begin{tabular}{c}
    \includegraphics[width=0.75\linewidth]{Number of Spectrogram.pdf} 
  \end{tabular}
  \label{STFT Spectrograms in Each Class}
\end{table}

\begin{itemize}
    \item The provided CSV files contain time-domain signals pertaining to the three primary faults. Specifically focusing on the rotor current and vibration signals, the data is organized into 5 .mat files, each representing 10 different torque loads sampled at a rate of 60 KHz. This compilation yields a total of 28,000 Short-Time Fourier Transform (STFT) images. Within the rotor dataset, with the exception of the healthy class, denoted as one, two, three, and four BRB, each category comprises 3000 spectral images. Similarly, the bearing dataset encompasses 67 files, each capturing data from 8 distinct sensor positions sampled at a consistent rate of 50 KHz. This amalgamation results in the creation of 10,000 STFT images. Among these, each fault category, excluding the healthy class, encompasses 3000 spectral images. Lastly, the stator dataset comprises 45 files sampled at differing frequencies: 25.6 KHz for vibration and 100 KHz for current. Consequently, this compilation produces 14,000 STFT images. The stator dataset encompasses classes representing both healthy states and specific fault conditions, such as ITSC and ICSC. STFT-based spectral spectrogram image distribution for each fault class is shown in Table\ref{STFT Spectrograms in Each Class}.

    \item The training set, comprising ${T_{r_1}, T_{r_2}, T_{r_3} \ldots, T_{r_n}}$ along with the validation set ${Vl_{1}, Vl_{2}, Vl_{3}, \ldots, Vl_{n}}$ is then subjected to various DL models.We employ six diverse DL architectures for fault diagnosis of IMs. This approach harnesses varied representations, enhancing feature extraction. By leveraging ensemble methods, we aim to boost robustness against model biases and improve prediction reliability. Integrating architectures with complementary strengths enables the enhancement of the understanding of the data. Furthermore, our strategy enhances generalization and mitigates overfitting. Each classifier, represented as ${C_{1}, C_{2}, \ldots, C_{6}}$ preserves the weights of its optimal model. Subsequently, four key multi-class evaluation measures ${V_{i}=Pre_{i}, Rec_{i}, F1_{i}, AUC_{i}}$ are derived. 
\subsection{Evaluation Measures}
These evaluation metrics encompass precision, recall, F1 score, and area under the ROC curve (AUC), which can be calculated utilizing equations \ref{eq8}, \ref{eq9}, \ref{eq10}, and \ref{eq11} respectively. 

A{c} is the proportion of accurately predicted instances relative to the total samples in the dataset can be derived using equation \ref{eq8}.
\begin{equation}
  \text{Accuracy} = \frac{\sum_{i=1}^{n} TP_{i}}{\text{Total number of instances}}
 \label{eq8}
\end{equation}
Precision assesses the accuracy of positive predictions by determining the ratio of correctly predicted positive instances to the total instances predicted as positive. This is calculated using the equation \ref{eq9}.
\begin{equation}
\text{Precision} = \frac{\sum_{i=1}^{C} \text{TP}_{i}}{\sum_{i=1}^{C} (\text{TP}_{i} + \text{FP}_{i})}
    \label{eq9}
\end{equation}
 Recall represents the proportion of accurately predicted positive instances among all labels belonging to the actual positive class to be calculated using equation \ref{eq10}.
\begin{equation}
\text{Recall} = \frac{\sum_{i=1}^{C} \text{TP}_{i}}{\sum_{i=1}^{C} (\text{TP}_{i} + \text{FN}_{i})}
 \label{eq10}
\end{equation}
The f-1 score, which balances precision and recall through the harmonic mean, is computed using equation \ref{eq11}.
\begin{equation}   
\text{F1-Score} = \frac{2 \times \text{Precision} \times \text{Recall}}{\text{Precision} + \text{Recall}}
 \label{eq11}
\end{equation}

    \item Following this step, the weights for each classifier are determined by employing equation \ref{eq12}.
    \begin{equation}
     w_i = \sum_{x \in v_i} \tanh(x) =\sum_{x \in v_i}\small\frac{\exp(x) - \exp(-x)}{\exp(x) + \exp(-x)}
     \label{eq12}
    \end{equation}
    \item After computing the weights of each classifier, we proceed to calculate the probabilities for every $j$th test sample of each $i$th classifier using the softmax function. Given the multi-class classification setup, softmax yields multiple probabilities for each $j$th sample of every $i$th classifier. The probabilities of each sample are calculated by equation \ref{eq13}

    \begin{equation}
    \sigma(\mathbf{T}_i) = \frac{e^{T_{ei}}}{\sum_{j=1}^{n} e^{T_{ei}}}, \quad \mathbf{T_e}_i = \{z_1, z_2, \ldots, z_n\}
    \label{eq13}
    \end{equation}
   
    \item The ensemble probability weight is determined by dividing the weighted sum of probabilities for the $j$th sample across all classes by the total weight assigned to the $j$th sample across all classes. Mathematically, this is expressed as:
    \begin{equation}
    \text{Ensemble Probability Weights} = \frac{\sum_{i} w_i \cdot P_{j}^{i}}{\sum_{i} w_i} 
    \label{eq14}
    \end{equation}
    \item  The final classification prediction is made by applying the double max operation on ensemble probability weights.
\end{itemize}   

\section{Results and Discussion}\label{sec5}
The following section outlines an extensive experimental examination conducted on multi-class classification datasets of bearings, rotors, and stators of IMs.The system configuration encompasses both hardware and software specifications. The hardware comprises an Intel(R) processor with a clock speed of 3.20 GHz and 32 CPU cores, alongside an NVIDIA RTX 3080 Ti GPU with 12GB of memory. Memory resources include 64 GB of DDR4 RAM, while storage is managed through a 256 GB SSD and a 1 TB HDD.
\begin{figure}[t]
        \includegraphics[width=1\linewidth]{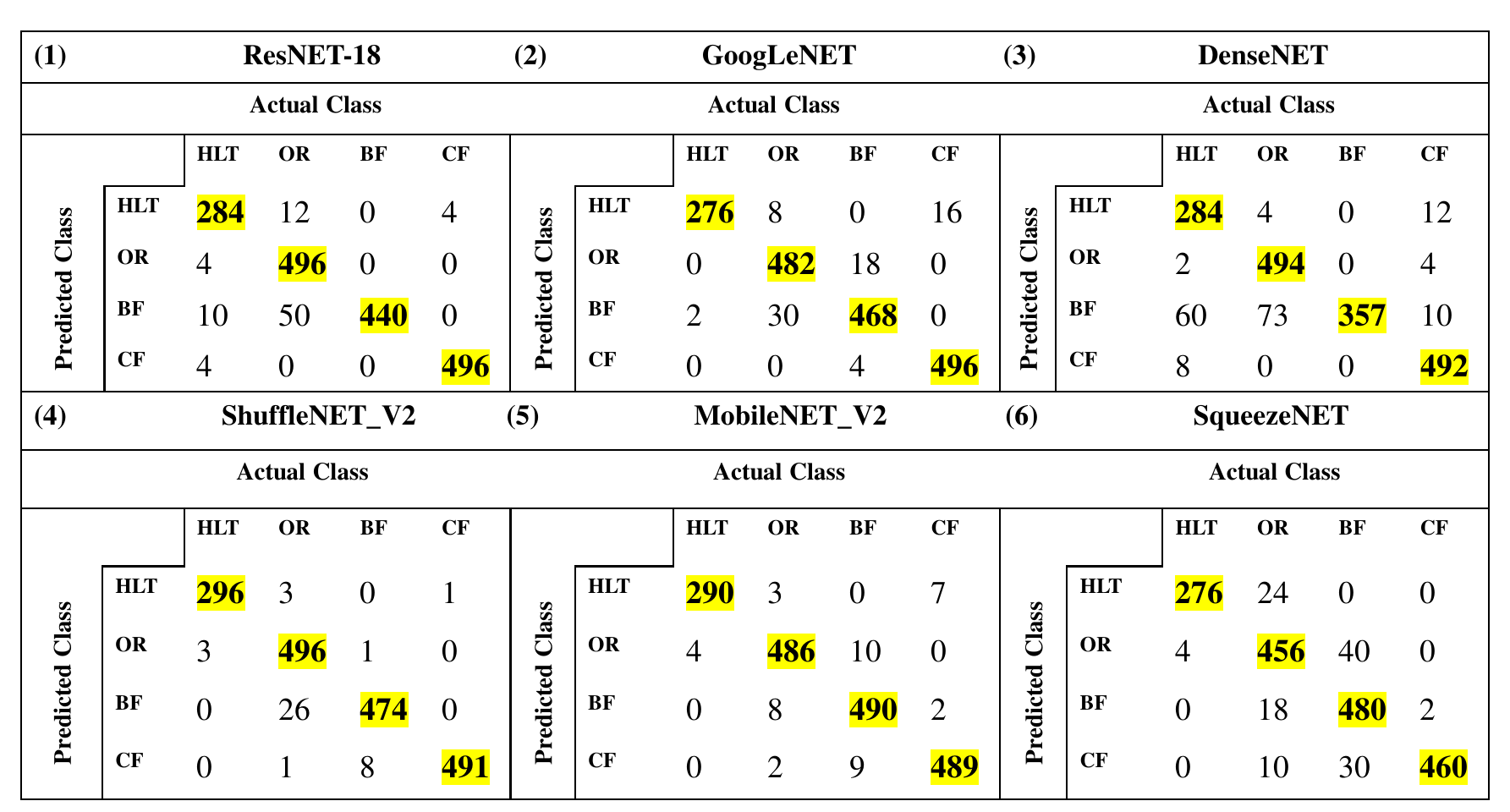} 
    \centering
    \caption{\centering Multi-Class Confusion Matrix of Fine-tune DL Models on Bearing Dataset}
    \label{fig:bearing cm}
\end{figure}
The initial experimental assessment is conducted on the bearing vibration dataset.
As indicated in Table \ref{STFT Spectrograms in Each Class}, this dataset comprises four classes: Healthy (HLT), ball fault (BF), outer race (OR), and cage fault (CF), with each class containing 1000, 3000, 3000, and 3000 STFT spectrograms, respectively. For training and evaluation purposes, we partitioned the dataset into 8200 images (82\%) for training and 1800 images (18\%) for testing. 
The detailed multi-class confusion matrix on the bearing dataset of each fine-tuned DL model is shown in figure \ref{fig:bearing cm}.

\begin{figure}[b]
  \centering    \includegraphics[width=0.7\linewidth]{Ensemble Bearing.pdf} 
  \caption{\centering Multi-Class Confusion Matrix of WPEDL Model on Bearing Dataset}
  \label{fig:bearing_vibration_cm}
\end{figure}

ResNET-18 demonstrates a 95.33\% accuracy, correctly predicting 1716 instances out of 1800, with 84 instances incorrectly classified. 
GoogleNET exhibits a 95.66\% accuracy rate, accurately identifying 1722 instances out of 1800, with 78 instances misclassified. DenseNET showcases a 90.33\% accuracy, correctly classifying 1627 instances out of 1800, with 173 instances erroneously labeled.

ShuffleNET-V2 achieves an impressive accuracy of 97.61\%, with 1775 correct classifications out of 1800, and only 25 misclassification. MobileNET-V2 attains a 97.5\% accuracy, accurately predicting 1755 instances out of 1800, with 45 instances incorrectly identified. 
The SqueezNet model achieves an accuracy of 92.85\%, accurately predicting 1672 instances out of 1800, with 128 misclassifications.

The maximum accuracy attained by the fine-tune DL model, ShuffleNet-V2, stands at 97.61\%. Figure \ref{fig:bearing_vibration_cm} illustrates the multi-class confusion matrix of our proposed WPEDL model applied to the bearing dataset. Notably, all instances within the healthy class are accurately predicted. 
In contrast, 498, 491, and 492 instances are correctly classified across the remaining classes, contributing to the overall enhancement of the final prediction accuracy to approximately 99.20\%.

Table \ref{tab2} compares the accuracies of different models from various studies in detecting bearing faults using vibration measurements. The models are evaluated for OR, BR, and CF detections and their overall accuracy. The WPEDL method demonstrates the highest overall accuracy among the referenced studies.

\begin{table*}[t]
\centering
\caption{Comparison of Our Approach with Other Techniques on Bearing Dataset}
\label{tab2}
\resizebox{\textwidth}{!}{%
\begin{tabular}{ccccccccccc}
\toprule
\textbf{Ref} & \multicolumn{4}{c}{\textbf{Bearing Faults}} & \textbf{Measurements} & \multicolumn{5}{c}{\textbf{Model Accuracy (\%)}} \\ \cmidrule(lr){2-5} \cmidrule(lr){7-11}
 & \textbf{HLT} & \textbf{OR} & \textbf{BR} & \textbf{CF} & & \textbf{HLT} & \textbf{OR} & \textbf{BR} & \textbf{CF} & \textbf{Overall} \\ \midrule
\cite{br1} & \cmark & \cmark & \cmark & \cmark & Vibration & - & - & - & - & 98.32 \\ \midrule
\cite{br2} & \cmark & \cmark & \cmark & \cmark & Vibration & 96.2 & 99.62 & 99.10 & 99.10 & 98.50 \\ \midrule
\cite{br3} & \cmark & \cmark & \cmark & \cmark & Vibration & - & 98.86 & 98.56 & 99.30 & 99.18 \\ \midrule
\cite{br4} & \cmark & \cmark & \cmark & \cmark & Vibration & 99.80 & 98.22 & 98.40 & 98.20 & 98.65 \\ \midrule
\textbf{WPEDL} & \cmark & \cmark & \cmark & \cmark & Vibration & \textbf{100} & \textbf{99.62} & \textbf{98.20} & \textbf{98.90} & \textbf{99.20} \\ \bottomrule
\end{tabular}}
\end{table*}

The second experimental evaluation involves analyzing the current and vibration signals from the stator dataset. This phase encompasses three distinct classes: HLT, ITSC, and ICSC, with a total of 14,000 STFT spectrograms. Half of these spectrograms originate from the current signal, while the remaining half derive from the vibration signal. The train-test split ratio and hyperparameters remain consistent with those utilized in our initial experiment.
\begin{figure}[http]
  \centering
  \begin{subfigure}{0.5\linewidth}
    \centering
    \includegraphics[width=\linewidth]{Stator Current.pdf}
    \caption{Confusion Matrix of DL models on Stator Current Dataset}
    \label{subtable:Stator_current}
  \end{subfigure}%
  \begin{subfigure}{0.5\linewidth}
    \centering
    \includegraphics[width=\linewidth]{Stator Vib.pdf}
    \caption{Confusion Matrix of DL models on Stator Vibration Dataset}
    \label{subtable:Stator_vibration}
  \end{subfigure}
   \caption{Comparison of Stator Current and Vibration Confusion Matrix }
  \label{fig:Stator_comparison}
\end{figure}
In figure \ref{subtable:Stator_current} and \ref{subtable:Stator_vibration}, the multi-class confusion matrices delineate the outcomes of fine-tuning DL models on the stator current and vibration datasets, correspondingly. While the highest correctly classified instances are attained on stator current and vibration data through the fine-tuned ResNET18 and DenseNET models at 96.3\% and 96.7\% respectively, it's crucial to note that other classifiers also play a role in the robustness of the final classification decision. This is evident in Figure \ref{subtable:Ensemble Stator_current} and \ref{subtable:Ensemble Stator_vibration}, where our WPE model augments the accuracy to 99\% and 99.30\% by bolstering the performance of the true positive class within the stator current and vibration datasets, respectively. This underscores the comprehensive approach employed, wherein various models enhance classification outcomes.
\begin{figure}[http]
  \begin{subfigure}{0.5\linewidth} 
    \includegraphics[width=\linewidth]{Ensemble Stator Current.pdf}
    \caption{\centering Confusion Matrix of WPEDL model on Stator Current Dataset}
    \label{subtable:Ensemble Stator_current}
  \end{subfigure}%
  \begin{subfigure}{0.5\linewidth} 
    \includegraphics[width=\linewidth]{Ensemble Stator Vib.pdf}
    \caption{\centering Confusion Matrix of WPEDL model on Stator Vibration Dataset}
    \label{subtable:Ensemble Stator_vibration}
  \end{subfigure}
  \caption{Comparison of Confusion Matrix of WPEDL model on Stator Current and Vibration Datasets}
  \label{fig:Ensemble_Stator_comparison}
\end{figure}

Table \ref{tab3} compares the accuracies of different methods from various studies in detecting stator faults using current and vibration measurements. The models are evaluated for HLT, ITSC, and ICSC detection. The WPEDL method demonstrates the highest accuracy among the referenced studies.

\begin{table*}[http]
\centering
\caption{Comparison of Our Approach with Other Techniques on Stator Dataset}
\label{tab3}
\label{table:rotor_fault_analysis}
\resizebox{\textwidth}{!}{%
\begin{tabular}{ccccccc}
\toprule
\textbf{Ref} & \multicolumn{3}{c}{\textbf{Fault Types}} & \multicolumn{1}{c}{\textbf{Measurements}} & \multicolumn{2}{c}{\textbf{Model Accuracy (\%)}} \\ \cmidrule{2-5} \cmidrule{6-7} 
                     & \textbf{HLT} & \textbf{ITSC} & \textbf{ICSC} & (\textbf{Current/Vibration)} & \textbf{Current} & \textbf{Vibration} \\
\midrule
\cite{st1}              & \checkmark & \checkmark & \checkmark & Both    & 43.20 & 83.00 \\
\cite{st2}              & \checkmark & \checkmark &  \checkmark         & Current & 98.2  & -     \\
\textbf{WPEDL}                  & \checkmark & \checkmark & \checkmark & \textbf{Both}  & \textbf{99.00} & \textbf{99.30} \\
\bottomrule
\end{tabular}
}
\end{table*}

The third experiment focuses on the rotor current and vibration signals, encompassing five distinct classes: HLT, BRB1, BRB2, BRB3, and BRB4. Notably, the healthy classes exhibit an imbalance, comprising 1000 and 3000 STFT spectrograms in the current and vibration datasets, respectively. The dataset comprises a total of 28,000 STFT images, with 13,000 originating from current signals and 15,000 from vibrational signals. The train-test split and hyperparameter configurations remain consistent with those utilized in our previous experiments.

\begin{figure}[http]
  \centering
  \begin{subfigure}{0.5\linewidth}
    \centering
    \includegraphics[width=\linewidth]{Rotor Current.pdf}
    \caption{Confusion Matrix of DL models on Rotor Current Dataset}
    \label{subtable:rotor_current}
  \end{subfigure}\hfill
  \begin{subfigure}{0.5\linewidth}
    \centering
    \includegraphics[width=\linewidth]{Rotor vib.pdf}
    \caption{Confusion Matrix of DL models on Rotor Vibration Dataset}
    \label{subtable:rotor_vibration}
  \end{subfigure}
  \caption{Comparison of Rotor Current and Vibration Confusion Matrix}
  \label{fig:rotor_comparison}
\end{figure}

In figures \ref{subtable:rotor_current} and \ref{subtable:rotor_vibration}, the multi-class confusion matrices provide insight into the outcomes of fine-tuning DL models on the rotor current and vibration datasets, respectively. While the highest rates of correctly classified instances are observed with the fine-tuned ShuffleNET-V2 and MobileNET-V2 models, achieving 98.70\% and 98.4\% accuracy, it's important to recognize the collective impact of all classifiers on the final classification decisions. This significance becomes apparent in Figures \ref{subtable:Ensemble Rotor_current} and \ref{subtable:Ensemble Rotor vibration}, where our WPEDL model significantly boosts accuracy to 99.60\% and 99.52\% by enhancing the performance of the true positive class within the rotor current and vibration datasets, respectively. This highlights the integrated nature of our approach, where various models work together synergistically to refine classification outcomes.    
\begin{figure}[http]
  \centering
  \begin{subfigure}{0.48\linewidth} 
    \centering
    \includegraphics[width=\linewidth]{Ensemble Rotor.pdf}
    \caption{Confusion Matrix of WPEDL model on Rotor Current Dataset}
    \label{subtable:Ensemble Rotor_current}
  \end{subfigure}\hfill
  \begin{subfigure}{0.48\linewidth} 
    \centering
    \includegraphics[width=\linewidth]{Ensemble Vibration.pdf}
    \caption{Confusion Matrix of WPEDL model on Rotor Vibration Dataset}
    \label{subtable:Ensemble Rotor vibration}
  \end{subfigure}
  \caption{Comparison of Rotor Current and Rotor Vibration on WPEDL model}
  \label{fig:Ensemble_Rotor_comparison}
\end{figure}

\begin{figure}[htbp]
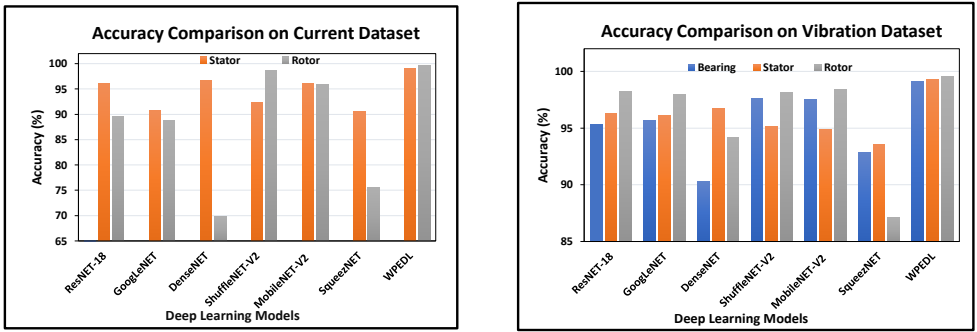

  \centering
  \begin{subfigure}{0.48\linewidth} 
    \centering
    \includegraphics[width=\linewidth]{Accuracy Comparison on Current Dataset.pdf}
    \caption{Accuracy Comparison on Current Dataset}
    \label{fig: Accuracy Comparison on Current Dataset}
  \end{subfigure}\hfill
  \begin{subfigure}{0.48\linewidth} 
    \centering
    \includegraphics[width=\linewidth]{Accuracy Comparison on Vibration Dataset.pdf}
    \caption{Accuracy Comparison on Vibration Dataset}
    \label{fig:Accuracy Comparison of DL models and WPEDL on Vibration Datasets of IMs}
  \end{subfigure}
  \caption{Accuracy Comparison of DL models and WPEDL on Current and Vibration Datasets of IMs}
  \label{fig:accuracy_comparison}
\end{figure}

Table \ref{tab4} compares the accuracies of different methods from various studies in detect-
ing rotor faults using current and vibration measurements. The models are evaluated
for BRB1, BRB2, BRB3, and BRB4 detection. The WPEDL method demonstrates the highest
accuracy among the referenced studies.
\begin{table*}[http]
\centering
\caption{Comparison of Our Approach with Other Techniques on Rotor Dataset}
\label{tab4}
\label{table:rotor_fault_analysis}
\resizebox{\textwidth}{!}{%
\begin{tabular}{cccccccc}
\toprule
\textbf{Ref} & & \textbf{Fault Type}
&  & \textbf{Measurement Type} & \textbf{Fault Analysis} & \multicolumn{2}{c}{\textbf{Model Accuracy (\%)}} \\ \cmidrule{2-5} \cmidrule{6-8} 
 &\textbf{1 BRB} & \textbf{2 BRB} & \textbf{Mult. BRB} & \textbf{(Curr./Vib.)} & \textbf{(Stdy./Trns.)} & \textbf{Curr.} & \textbf{Vib.} \\ \midrule
\cite{rc1} & \checkmark & \checkmark & \xmark & Vib. & Stdy & \xmark & 97.78 \\ 
\cite{rc2} & \checkmark & \checkmark & \checkmark & Vib. & Both & \xmark & 97.67 \\ 
\cite{rc3} & \checkmark & \checkmark & \xmark & Curr. & Stdy & 95.8 & \xmark \\ 
\cite{rc4} & \checkmark & \checkmark & \xmark & Curr. & Stdy & 99 & \xmark \\ 
\textbf{WPEDL} & \checkmark & \checkmark & \checkmark & \textbf{Both} & \textbf{Both} & \textbf{99.60} & \textbf{99.52} \\ \bottomrule
\end{tabular}
}
\end{table*}
The comparison of accuracies for each model on the current and vibration datasets of bearing, rotor, and stator is shown in figure \ref{fig: Accuracy Comparison on Current Dataset} and figure \ref{fig:Accuracy Comparison of DL models and WPEDL on Vibration Datasets of IMs}, respectively.

\begin{figure}[t]
  \centering
   \begin{tabular}{c}
    \includegraphics[width=0.99\linewidth]{Ensemble Combined Prediction.pdf} 
  \end{tabular}
   \caption{Confusion Matrix of WPEDL model on Combined Faults Dataset}
  \label{fig:Ensemble Combined_cm}
\end{figure}
The culmination of our research lies in the final experiment, where we scrutinize the robustness of our proposed WBELD model across a combined dataset integrating both current and vibration data on all faults. We aim to probe its efficacy in navigating the intricate landscape of fault detection. Within this comprehensive study, the dataset encompasses 52,000 STFT spectrograms, presenting various fault scenarios and healthy states.

Delving into the dataset, we find the healthy class consisting 7,000 STFT images, partitioned with 5,000 for training and 2,000 reserved for model testing. Similarly, the ITSC and ICSC categories each contribute 6,000 samples, divided into training and testing subsets of 5,000 and 1,000, respectively. The rotor classes—BRB1, BRB2, BRB3, and BRB4—follow suit with their own 6,000 samples, allocated for training and testing in proportions mirroring the previous categories. Additionally, the BF and CF classes make their presence known with 3,000 samples each, thoughtfully divided to ensure a robust training-testing balance.

In this dataset, we employ a weighted probability methodology to measure key performance metrics including precision, recall, F1 score, and AUC score. This approach allows us to paint a realistic picture of the WPELD model's effectiveness and compare it against the fine-tuned DL models in its ability to navigate the multifaceted challenges inherent in fault detection tasks.

In continuation with our investigation, figure \ref{fig:Ensemble Combined_cm} displays the ensemble classification confusion matrix, offering insights into the performance of our model on multi-class IMs faults. The diagonal elements of this matrix represent instances correctly classified as true positives by our model. Impressively, out of 9500 test images, a total of 9389 predictions are accurately classified. 
Table \ref{tab5} presents the performance metrics of the WPEDL framework evaluated through various ablation experiments. Each row corresponds to a different combination of model weights (\(w1\) through \(w6\)) used in the experiments. The classification report includes accuracy, precision, recall, F1-score, and AUC. The final row shows the best performance with all weights combined, achieving the highest metrics across all categories: accuracy (98.89\%), precision (98.25\%), recall (98.28\%), F1-score (98.33\%), and AUC (98.99\%).

\begin{table*}[http]
\centering
\caption{Performance Metrics of WPEDL Across Various Ablation Experiments}
\label{tab5}
\resizebox{\textwidth}{!}{%
\begin{tabular}{ccccccc ccccc}
        \hline
        \multicolumn{6}{c}{\textbf{Model Weights}} & \multicolumn{5}{c}{\textbf{Classification Report}} \\
        \textit{\textbf{w1}} & \textit{\textbf{w2}} & \textit{\textbf{w3}} & \textit{\textbf{w4}} & \textit{\textbf{w5}} & \textit{\textbf{w6}} & \textbf{Accuracy} & \textbf{Precision} & \textbf{Recall} & \textbf{F1-Score} & \textbf{AUC} \\
        \hline
        & & & & & $\checkmark$ & 91.78 & 93.50 & 91.78 & 91.86 & 94.30 \\
        & $\checkmark$ & & & & & 94.32 & 95.03 & 94.3 & 94.21 & 96.52 \\
        & & $\checkmark$ & & & & 96.71 & 96.59 & 96.71 & 96.71 & 96.72 \\
        $\checkmark$ & $\checkmark$ & & & & & 97.11 & 97.29 & 97.11 & 97.01 & 97.61 \\
        $\checkmark$ & $\checkmark$ & & $\checkmark$ & & & 97.19 & 97.14 & 97.05 & 97.09 & 97.28 \\
        $\checkmark$ & $\checkmark$ & $\checkmark$ & & & & 97.35 & 97.23 & 97.31 & 97.01 & 97.71 \\
        $\checkmark$ & $\checkmark$ & $\checkmark$ & & $\checkmark$ & & 97.77 & 97.69 & 97.72 & 97.81 & 97.91 \\
        $\checkmark$ & $\checkmark$ & $\checkmark$ & $\checkmark$ & $\checkmark$ & $\checkmark$ & \textbf{98.89} & \textbf{98.25} & \textbf{98.28} & \textbf{98.33} & \textbf{98.99} \\
        \hline
    \end{tabular}}
\end{table*}

\begin{figure}[http]
  \centering
  \begin{tabular}{c}
    \includegraphics[width=0.6\linewidth]{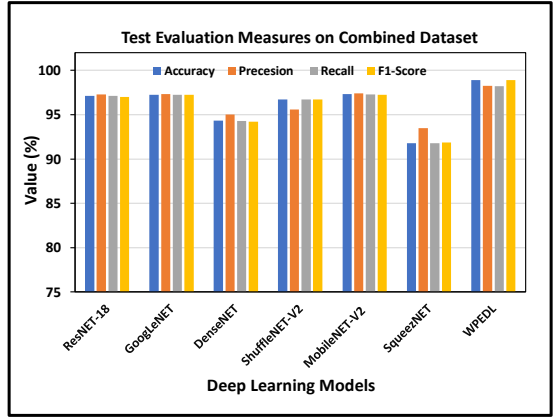} 
  \end{tabular}
   \caption{Accuracy Comparison of DL models and WPEDL on Combined Datasets of IMs}
  \label{fig:Accuracy Comparison of DL models and WPEDL on Combined Datasets of IMs}
\end{figure}

The barplot comparison of different evaluation measures i.e., accuracy, precision, recall, and f1-score, for DL models and WPEDL model on the combined datasets of bearing, rotor, and stator is shown in figure \ref{fig:Accuracy Comparison of DL models and WPEDL on Combined Datasets of IMs}.

\section{Conclusion}\label{sec6}

This research article introduced the WPEDL technique as a robust approach for early-stage fault diagnosis in IMs. By leveraging high-dimensional data extracted from vibration and current features, WPEDL demonstrates superior efficacy in diagnosing various fault types encountered in IMs, including bearing, rotor, and stator faults. A comparison with conventional deep learning models highlights WPEDL's superior performance in fault diagnosis. Furthermore, WPEDL achieves high accuracies across different fault types, with accuracies of 99.05\% for bearing faults, 99.10\% and 99.50\% for rotor current and vibration datasets, and 99.60\% and 99.52\% for stator current and vibration datasets, respectively. Evaluation of WPEDL's robustness through tests on a combined dataset, which correctly classified 98.89\% of test cases, further solidifies its utility in industrial settings. These findings suggest that WPEDL holds significant promise for enhancing industrial operational efficiency and reliability by facilitating early fault detection in IMs.

\bibliography{sn-bibliography}

\end{document}